\documentclass{article}

\usepackage[preprint]{neurips_2022}

\usepackage[utf8]{inputenc} 
\usepackage[T1]{fontenc}    
\usepackage{hyperref}       
\usepackage{url}            
\usepackage{subcaption}
\usepackage{booktabs}       
\usepackage{amsfonts}       
\usepackage{nicefrac}       
\usepackage{microtype}      
\usepackage{xcolor}         
\usepackage{graphicx}
\usepackage{multirow}
\usepackage{siunitx}
\usepackage{array}
\usepackage{bm}
\usepackage{todonotes}
\usepackage{amssymb}
\usepackage{pifont}
\newcommand{\cmark}{\ding{51}}%
\newcommand{\xmark}{\ding{55}}%

\title{Self-supervised Learning for Human Activity Recognition Using 700,000 Person-days of \\ Wearable Data}

\author{%
Hang Yuan$^{1,2,4}$* \ \ Shing Chan$^{1,2}$* \thanks{Equal contributions. Correspondence to aiden.doherty@ndph.ox.ac.uk. The code and model can be downloaded from \url{https://github.com/OxWearables/ssl-wearables}.}\ \ Andrew P. Creagh$^{2,3}$\ \  Catherine Tong$^4$ \\  \textbf{Aidan Acquah} $^{2,3}$ \ \textbf{David A. Clifton}$^3$\ \   \textbf{Aiden Doherty}$^{1,2}$  \\
${}^1$ Nuffield Department of Population Health, University of Oxford \\ 
${}^2$ Big Data Institute, University of Oxford \\ 
${}^3$ Department of Engineering Science, University of Oxford \\ 
${}^4$ Department of Computer Science, University of Oxford  \\ 
}

\begin{document}

\maketitle

\begin{abstract}
Advances in deep learning for human activity recognition have been relatively limited due to the lack of large labelled datasets. In this study, we leverage self-supervised learning techniques on the UK-Biobank activity tracker dataset--the largest of its kind to date--containing more than 700,000 person-days of unlabelled wearable sensor data. Our resulting activity recognition model consistently outperformed strong baselines across seven benchmark datasets, with an F1 relative improvement of 2.5\%-100\% (median 18.4\%), the largest improvements occurring in the smaller datasets. In contrast to previous studies, our results generalise across external datasets, devices, eand environments. Our open-source model will help researchers and developers to build customisable and generalisable activity classifiers with high performance.
\end{abstract}

\section{Introduction}

Current human activity recognition (HAR) models typically rely on manual feature engineering~\citep{twomey2018comprehensive, haresamudram2019role} partly due to the very limited size of existing labelled datasets. This small data issue caps the effectiveness of data-hungry deep learning methods. In general, obtaining \emph{labelled} data is labour intensive, but it is especially so for HAR data because one would need to annotate the corresponding video stream for the ground truth. On the other hand, collecting large-scale \emph{unlabelled} HAR data is highly feasible, as evidenced by projects such as the UK-Biobank (UKB)~\citep{doherty2017large} and NHANES~\citep{belcher2021us}. This prompts the use of self-supervised learning (SSL) methods to leverage unlabelled data in a similar manner as language models~\citep{mikolov2013efficient, devlin2018bert, radford2018improving, radford2019language} and vision models~\citep{doersch2015unsupervised, zhang2016colorful, noroozi2016unsupervised, Wei2018, he2020momentum, chen2020simple}.  Recent studies explored the utility of SSL for HAR~\citep{saeed2019multi, tang2020exploring}, but these still relied on small-scale laboratory-style datasets, hence the full potential of SSL-HAR remains unknown.

\begin{figure}[t]
		\centering
		\includegraphics[width=.85\linewidth]{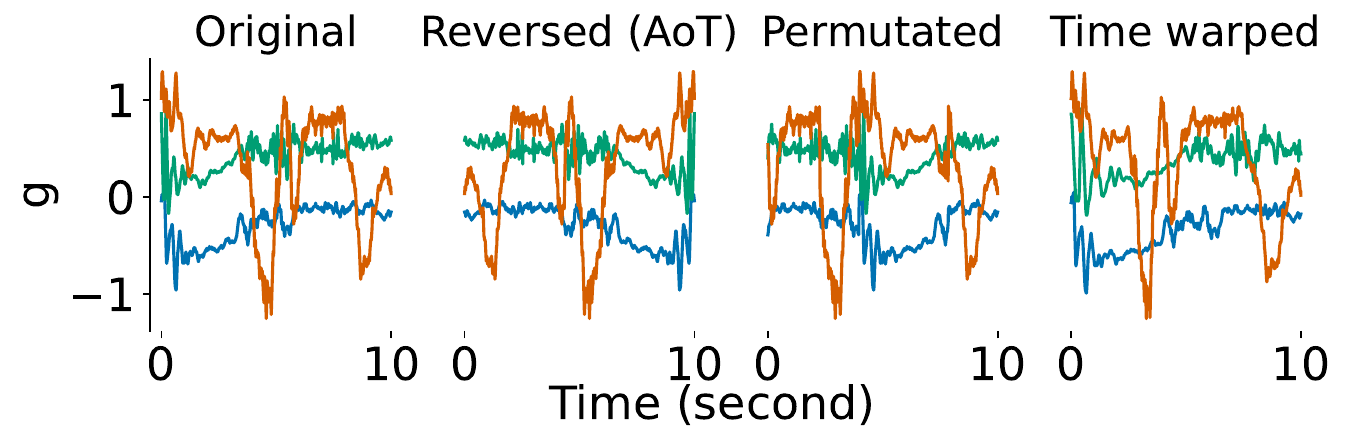}
		 \caption{A ten-second window of tri-axial accelerometer data when a person is ironing in its original form and three corresponding transformations: reversed, permuted, and time-warped.}
        \label{fig:mtlVisu}
\end{figure}

In this paper, we investigate how learning three simple self-supervised learning tasks independently and jointly could facilitate HAR across a variety of environments using the UKB dataset, which contains terabytes of wearable sensor data collected in the real world. To fully appreciate the benefit of SSL in HAR, we chose the tasks that would prioritise the temporal dependencies of human motion, namely, \emph{arrow of time} (AoT), \emph{permutation} and  \emph{time-warping} (TW)~\citep{TerryUm_ICMI2017}. Figure~\ref{fig:mtlVisu} illustrates an example of these transformations for a sequence of tri-axial accelerometer data. We show for the first time that multi-task SSL can train a HAR model that generalises well across seven external datasets that differ in activity classes, devices, populations and recording environments. In contrast to previous works, we provide a more realistic evaluation of the utility of SSL-HAR by factoring in common issues seen in the practical use cases of pre-trained models such as domain shift and task shift~\citep{quinonero2008dataset}. Our publicly available models will enable the research community to build high-performing activity recognition models even in a resource-restricted environment.

\section{Related work}
Activity recognition from on-body inertial sensors is a fundamental problem in wearable computing. HAR models are usually built following the Activity Recognition Chain \citep{bulling2014tutorial}, which describes the pre-processing and segmentation of raw time series into fixed-length frames, from which features are extracted for classification. Early work in HAR found some success in tree-based approaches with hand-crafted features \citep{hammerla2013preserving}. 
More recent works focused on translating advances in deep learning into HAR~\citep{yang2015deep, hammerla2016deep, ronao2016human} but were limited by small dataset sizes.
Indeed, some studies found that earlier approaches, such as random forests, tended to generalise better than deep learning models in certain small data regimes~\citep{willetts2018statistical, kwon2020imutube}.

Recent studies explored the use of SSL for HAR, such as forecasting~\citep{taghanaki2020self}, masked reconstruction~\citep{haresamudram2020masked}, contrastive learning~\citep{haresamudram2020contrastive, tang2020exploring}, and multi-task SSL~\citep{saeed2019multi, tang2021selfhar}. 
However, these studies were still limited by small laboratory-style datasets ($n< 10{\small,}000$), which confounded their findings.
In comparison, we leverage the largest ever collection of wrist-worn raw accelerometer data ($n > 6\;\textrm{billion}$) collected in the real world. Aside from being several orders of magnitude larger, this dataset is vastly more diverse as it contains hundreds, if not thousands, of natural human activities -- a crucial aspect regarding the input distribution and generalisation. Therefore, we aim to provide a more definitive investigation of the utility of SSL for HAR.

\section{Methods}
\newcolumntype{A}{>{\centering}p{2cm}}
\newcolumntype{B}{>{\centering}p{1.05cm}}
\newcolumntype{C}{>{\centering}p{2.05cm}}
\newcolumntype{D}{>={\centering}p{2.0cm}}

\begin{table}[t]
{\footnotesize
	\caption{Wrist-worn accelerometer datasets used to evaluate the utility of self-supervised learning for human activity recognition tasks}
	\label{tab:dataset}
	\vskip 0.15in
	\begin{center}
			\begin{tabular}{ABBBCc}
				\toprule
				Dataset & \#Subjects & \#Samples &  \#Classes &  Environment & References \\
				\midrule
				UK-Biobank & $\sim$100K  & 6 B & Unlabelled & Free-living & ~\citet{doherty2017large} \\
				Capture-24 & 152 & 573K   & 4  & Free-living & ~\citet{willetts2018statistical} \\
				Rowlands &  55  & 36K & 13  & Lab & ~\citet{eslinger2011validation}  \\
				WISDM    & 46  &  28K & 18  & Semi free-living & ~\citet{weiss2019smartphone}  \\
                REALWORLD    & 14  &  12K & 8  & Lab & ~\citet{sztyler2016body} \\
            	Opportunity &  4  & 3.9K & 4  & Semi free-living & ~\citet{roggen2010collecting}\\
				PAMAP2    & 8  &  2.9K & 8  & Lab & ~\citet{reiss2012introducing}  \\
				ADL &  7  & 0.6K & 5  & Lab & ~\citet{bruno2013analysis} \\
				\bottomrule
			\end{tabular}
	\end{center}
	\vskip -0.1in
}
\end{table}

We used tri-axial accelerometer data from wrist-worn activity trackers, which record acceleration on three orthogonal axes at a high sampling rate (e.g. \SI{100}{\hertz}). The main benefit of wrist-worn activity trackers is their high user compliance, resulting in days, if not weeks, of continuous recordings. Following~\citet{bulling2014tutorial}, we split the signals into windows of equal duration, effectively treating them as independent inputs to the HAR models. We can then label each window with an activity class. Throughout this study, we linearly resampled all data to \SI{30}{\hertz} resolution and used ten-second-long windows to compare the downstream benchmarks fairly. The \SI{30}{\hertz} sampling rate was used because most human activities have a frequency less than~\SI{10}{\hertz}. We used a sampling rate that is higher than the presumed Nyquist rate (\SI{20}{\hertz}) to ensure that we did not lose any useful signal.

\subsection{Datasets} \label{sec:datasets}
Our multi-task SSL training relied on the \emph{unlabelled} \textit{UKB} dataset, which contains roughly 700,000 person-days of free-living activity data (>100,000 participants, seven days of wear). The free-living aspect is important because the data can contain all sorts of activities, as opposed to lab data which are constrained to scripted activities only. The UKB data (project ref Anonymous) is covered by ethical approval from the NHS National Research Ethics.

For the subsequent activity recognition benchmarks, we considered seven external \emph{labelled} datasets that vary in size (600 to 600,000 samples), activity classes (4 to 18 classes), devices used (five different brands), and collection protocol (free-living, scripted, and lab settings). See Table~\ref{tab:dataset} for additional details. Three datasets had license information, and four datasets had explained informed consent information (Appx. Table~\ref{tab:extraInfo}). We removed the classes that were not present in all the subjects in small datasets with less than 10 individuals during data cleaning.

Even though we reused existing datasets, we made our best effort to enumerate the license and consent information for all the included datasets, as our data involved human subjects. We observed that many open benchmark datasets that we used did not have suitable licencing or consent information, possibly due to the lack of data governance awareness at the time of the study.

\subsection{Multi-task self-supervised-learning}
We considered three self-supervised tasks from ~\citet{saeed2019multi}, which were first used by ~\citep{TerryUm_ICMI2017} as data augmentation techniques. 

Arrow of time (\textbf{AoT})  flips the signal along the time axis, effectively playing the signal in reverse. 

\textbf{Permutation} breaks the signal into chunks and shuffles them. We set the number of chunks to four and the minimum length of each chunk to at least 10 timestamps. 

Time warping (\textbf{TW}) stretches and compresses arbitrary segments of the signal, effectively slowing down and speeding up the signal randomly.

Following~\citet{saeed2019multi}, we treated each of the tasks as a binary problem predicting whether a transformation has been applied. In the multi-task learning (MTL) setting, not all the tasks might benefit HAR when trained jointly, so we assessed how different task combinations could influence the downstream performance (Section~\ref{sec:mtl_1k}).
%
%
We computed the cross-entropy loss for each task and weighed all the tasks equally in the loss calculation.

\paragraph{Weighted sampling}
Motion data collected in the real world contains large portions of low movement periods that are less informative (Appx. Figure~\ref{fig:sd}), which is an issue for our SSL tasks as static signals remain virtually unchanged after the transformations.
We found it crucial to perform weighted sampling for improved training stability and convergence: During training, we sample the data windows in proportion to their standard deviation so as to give more weight to high-movement periods.

\subsection{Network training} \label{sec:networkTrain}
We adapted a ResNet-V2 with 18 layers and 1D convolutions \citep{he2016identity} for the main trunk (feature extractor), totalling 10M parameters. The learned feature vector was of size 1024. All the tasks shared the same feature extractor. Then, we attached a softmax layer for each of the self-supervised tasks.  In the downstream evaluation, we added a fully-connected (FC) layer of size 512 in between the feature extractor and softmax readout. The network structure was fixed for all the downstream evaluations. 

For SSL, we load up to four subjects from the \textit{UKB} at each iteration. For each subject, we first sampled one day out of the week-long data, from which we again sampled 1500 10-second windows to make up a training batch. Self-supervised transformations were then applied to the batch of data. Since the axis orientation differs between device manufacturers, we used random axis swaps and rotations to augment the training data to embed this invariance into our models. For optimisation, we used Adam~\citep{kingma2014adam} with a learning rate of 1e-3. To account for large batch sizes, $1500 \times 4 = 6000$, we applied linear scaling for the learning rate with 5 epochs as burn-in \citep{goyal2017accurate}.  We distributed the network training over four Tesla V100-SXM2 with 32GB of memory. It took about 420 GPU hours to train the MTL model (about 20 epochs). We used an 8:2 ratio for the train/test split for all the self-supervised experiments. For fine-tuning, we used the same training setup as the pre-training where possible except for the batch size, which was re-adjusted depending on the size of each dataset.

\subsection{Evaluation -- human activity recognition}
To evaluate the downstream (HAR) performance, we used held-one-subject-out cross-validation for the datasets that had $ < 10$ subjects. We additionally removed activity classes not done by all the subjects in these small datasets. For datasets with $\ge 10$ subjects, we used five-fold subject-wise cross-validation instead. Each cross-validation had a 7:1:2 split ratio for train/validation/test sets. We used early-stopping with a patience of five to avoid over-fitting. 

After the network was trained on the \textit{UKB} using $\sim$100,000 participants, we further fine-tuned the network on the seven labelled downstream datasets to perform human activity detection using two approaches: (1) fine-tuning all the layers (2) freezing the trunk (feature extractor) and fine-tuning only the FC layers in the end.
We also report the model performance for a network of the same architecture but fully trained from scratch, and a strong random forest model with tried-and-tested time series features, which has often been neglected in baseline model comparisons~\citep{zhang2012physical,mannini2013activity,ellis2016hip,willetts2018statistical}. See Appx.~\ref{appx:feature} for the list of features used.

In addition,  a \textbf{shared implementation} was introduced for our network training, model evaluation and preprocessing. Differences in experiment setup such as training rates, regularisation and data augmentation can lead to inconsistent results~\citep{oliver2018realistic}. A unified evaluation framework would ensure a fair comparison between different baseline models. Our evaluation framework contrasts with previous work, where there is no fixed evaluation protocol across the benchmark datasets, making it hard to compare model performance with the current state-of-the-art. The results produced in our paper would serve as the baseline for future HAR research.

\paragraph{Transfer learning}
Pre-training on a larger labelled-dataset and fine-tuning on a smaller dataset is a common technique in practical application that has been under-reported as a baseline for SSL. The success of transfer learning, however, depends on how similar the source and target datasets are.  Hence, we included experiments using the two largest labelled datasets, \textit{Capture-24} and \textit{Rowlands} for pre-training, which were then fine-tuned on other labelled datasets.   

\paragraph{The benefits of data volume}
In the ablation studies, we investigated how the downstream performance differ on two axes, the amount of labelled data and the amount of unlabelled data. Concretely, we gradually increase the number of labelled subjects in both \textit{Capture-24} and \textit{Rowlands} in the downstream evaluation to assess whether our pre-trained model can still do well in a limited-data regime. In terms of unlabelled data, we experimented with pre-training that had 100 to 100,0000 participants with one order of magnitude increment. We also varied the amount of unlabelled data per subject from .25 to 1 using 10,000 participants. A data ratio of .25 means that if one day of data per subject was used previously, then only six hours of data per subject would now be used for training. Investigating how unlabelled data influences downstream performance will guide how much data one needs to have to obtain an effective SSL model for HAR.

\subsection{Understanding network representation}

\paragraph{Contextualising layer-wise relevance propagation}
 We applied layer-wise relevance propagation (LRP) to visually investigate the signal characteristics relevant for detecting the pretext tasks~\citep{montavon2019lrp, creagh2021interpretable}. It is inherently more difficult to visually interpret attribution heatmaps generated through Explainable AI (XAI) frameworks on time-series signals. To overcome this lack of visual ground truth, we devised a set of simple contextual experiments to evaluate our LRP attribution results. Using the same accelerometer as the \textit{UKB}, we recorded a participant performing two activities under video observation: (1) low intensity scripted (hand-shaking) and (2) high intensity unscripted (playing tennis). We acquired a ground truth (the context) for the accelerometer activity through the time-synced video observations, enabling a better visual interpretation of the sensor-based characteristics attributed as relevant for detecting different pretexts. Holistic interpretations were formed based on visualising the raw sensor signal, its analogues time-frequency representation through continuous wavelet transform (CWT) scalograms \citep{RN701}, as well as the time- and pretext task-localised LRP relevance scores, all with respect to observing the concurrent video recordings. Details on the XAI contextual LRP (cLRP) framework are described in Appx. \ref{sec:appendix:methods:xai}.

\section{Results}
\subsection{Weighted single task training}\label{sec:weighted_sampling}
\begin{figure}[t]
	\vskip 0.2in
	\begin{center}
		\centerline{\includegraphics[width=.62\columnwidth]{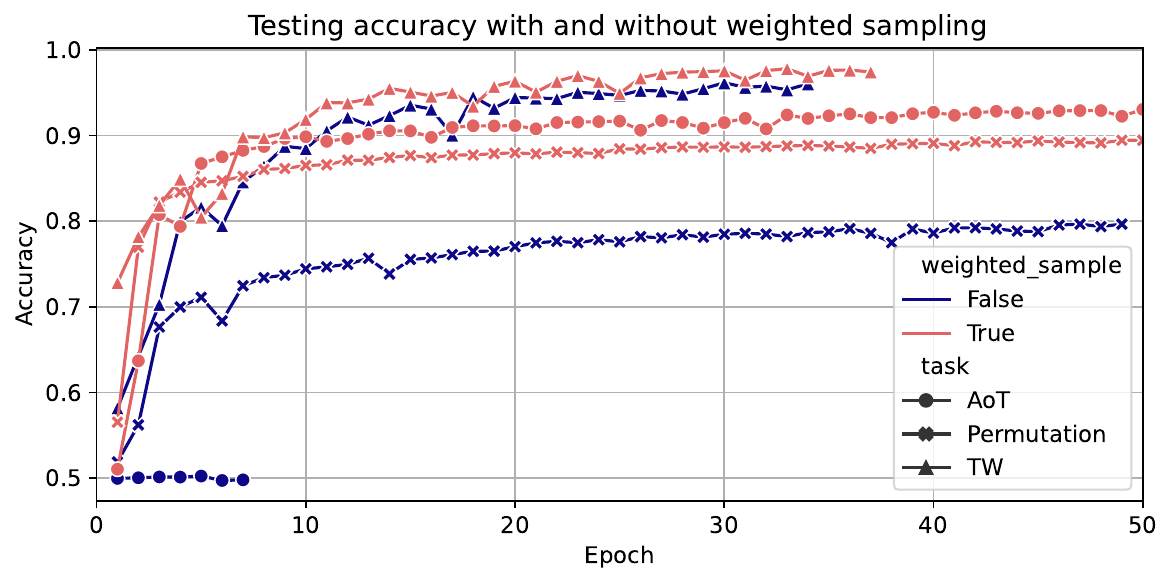}}
		\caption{Accuracy test curves for training four self-supervised tasks individually using 1000 subjects from the UK-Biobank with and without weighted sampling. The patience for early-stopping was five.}
		\label{fig:weighted_sample}
	\end{center}
	\vskip -0.2in
\end{figure}
When training individual pretext tasks, we found that without weighted sampling, all the tasks had worse convergence behaviour (Figure~\ref{fig:weighted_sample}). The performance degradation was most pronounced for the AoT and permutation. The test performance for the AoT stayed at the random chance level, and the test performance for permutation dropped roughly 10 percentage points without weighted sampling.

\subsection{Multi-task self-supervised learning} \label{sec:mtl_1k}
\begin{table}[t]
{\small
	\caption{Downstream human activity recognition performance (subject-wise F1 ($\pm$SD)) for different self-supervised task combinations using 1,000 UK-Biobank participants. $N$ is the number of samples.} 
	\label{tab:mtl}
	\vskip 0.15in
	\begin{center}
			\begin{tabular}{cccccc}
				\toprule
				\multirow{2}{*}{AoT}  &   \multirow{2}{*}{Permutation} &   \multirow{2}{*}{TW}  & Capture-24    &  Rowlands  & Opportunity \\
                & &  &  n=573K & n=36k & n=3.9K   \\ 
				\toprule
                Single task \\
                \cmark & \xmark  &  \xmark &   .671 $ \pm $ .094 & .565 $ \pm $ .120   &  .582 $ \pm $ .054 \\
                \xmark & \cmark  &  \xmark  & .721 $ \pm $ .093 & .783 $ \pm $ .099  & \textbf{.588 $ \pm $ .076}   \\
                \xmark & \xmark  &  \cmark &  .715 $ \pm $ .093 & .776 $ \pm $ .110  &  .584 $ \pm $ .064  \\
                Multi-task \\
    			 \xmark & \cmark  &  \cmark	 & .714 $ \pm $ .094 & .755 $ \pm $  .103 &  .587 $ \pm $ .070 \\
				 \cmark & \xmark  &  \cmark	 & \textbf{.719 $ \pm $ .094} & .762 $ \pm $  .102 & .530 $ \pm $ .071  \\
    			 \cmark  & \cmark  &  \xmark   	 & .718 $ \pm $ .092 &  \textbf{.781 $ \pm $  .101} & .502 $ \pm $ .081 \\
				 \cmark & \cmark  &  \cmark & .718 $ \pm $ .095 & .770 $ \pm $  .102  & .482 $ \pm $ .078 \\
				\bottomrule
			\end{tabular}
	\end{center}
	\vskip -0.1in
}
\end{table}

 To investigate how different SSL configurations perform in three downstream datasets, we picked one large (\textit{Capture-24}), medium (\textit{Rowlands}) and small (\textit{Opportunity}) dataset for evaluation. We trained different tasks both individually and jointly using 1000 subjects from the \textit{UKB}, then we fine-tuned the models on the subsequent HAR benchmarks (Table~\ref{tab:mtl}).
 
The differences between different SSL combinations on large datasets (\textit{Capture-24} and \textit{Rowlands}) was smaller than that of the smaller dataset (\textit{Opportunity}). There was no clear best performing configuration, and thus, for ease of comparison, we chose to use all tasks in pre-training for the remaining experiments. In addition, training more tasks together might yield the most general representation for different downstream datasets.

\subsection{Downstream performance - human activity recognition}
\begingroup
\setlength{\tabcolsep}{4pt}
\begin{table}[t]
{\small
	\caption{Subject-wise F1 and Kappa ($\kappa$) for downstream human activity recognition tasks (mean $\pm$ SD) using 100,000 participants for pre-training. The relative improvement compares the performance between the model that is trained from scratch and fine-tuning using all the layers.} 
	\label{tab:benchmark}
	\vskip 0.15in
	\begin{center}
			\begin{tabular}{ cc c| c cc | c}
				\toprule
              \multirow{3}{*}{Data}	 & & \multirow{3}{*}{Random forest} &  \multicolumn{3}{c|}{ResNet} & \multirow{3}{*}{Improvement}  \\
               \cline{4-6}
				 &   &	 &  \multirow{2}{*}{Trained from scratch}	 & \multicolumn{2}{c|}{Fine-tune self-supervised} \\
				&  &  &  &  After ConV layers & All layers    \\
				\toprule
				
				\multirow{2}{*}{Capture-24} & F1 & .694 $ \pm $ .099 & .708 $ \pm $ .094 & .723 $ \pm $ .097 & \textbf{.726 $ \pm $ .093} & 2.5 \%  \\
				
				& $\kappa$ & .683 $ \pm $ .101 & .703 $ \pm $ .092 & .718 $ \pm $ .090 & \textbf{.737 $ \pm $ .087} & 4.8\%\\%
				
				\midrule
				\multirow{2}{*}{Rowlands}& F1  & .700 $ \pm $ .090 & .696 $ \pm $ .106 & .724 $ \pm $ .081 & \textbf{.796 $ \pm $ .093} & 14.4\% \\
				& $\kappa$ & .830 $ \pm $ .086 & .810 $ \pm $ .098 & .850 $ \pm $ .062 & \textbf{.874 $ \pm $ .073} & 7.9\%\\%
		
				\midrule
				\multirow{2}{*}{WISDM}& F1  & .711 $ \pm $ .149 & .684 $ \pm $ .123 & .759 $ \pm $ .121 & \textbf{.810 $ \pm $ .127} & 18.4\%\\
				& $\kappa$ & .715 $ \pm $ .153 & .685 $ \pm $ .124 & .758 $ \pm $ .121 & \textbf{.809 $ \pm $ .126} & 18.1\% \\%

				\midrule
				\multirow{2}{*}{REALWORLD} & F1  & .731 $ \pm $ .119 & .705 $ \pm $ .062 & .764 $ \pm $ .052 & \textbf{.792 $ \pm $ .075} &  12.3\% \\
				& $\kappa$ &.680 $ \pm $ .142 & .638 $ \pm $ .079 & .703 $ \pm $ .063 & \textbf{.739 $ \pm $ .086} & 15.8\%\\%

				\midrule
				\multirow{2}{*}{Opportunity}	& F1 &.416 $ \pm $ .185 & .383 $ \pm $ .124 & .570 $ \pm $ .078 & \textbf{.595 $ \pm $ .085} & 55.4\% \\
				& $\kappa$ & .318 $ \pm $ .206 & .238 $ \pm $ .154 & .435 $ \pm $ .092 & \textbf{.471 $ \pm $ .104}  & 97.9\%\\%
				
				\midrule
				\multirow{2}{*}{PAMAP2}	& F1 & .753 $ \pm $ .093 & .605 $ \pm $ .086 & .725 $ \pm $ .054 & \textbf{.789 $ \pm $ .054}& 30.4\%	 \\
				& $\kappa$ &.744 $ \pm $ .101 & .596 $ \pm $ .086 & .717 $ \pm $ .057 & \textbf{.769 $ \pm $ .059} & 29.0 \%\\%

				\midrule
				\multirow{2}{*}{ADL}	& F1 & .764 $ \pm $ .180 & .414 $ \pm $ .179 & .645 $ \pm $ .107 & \textbf{.829 $ \pm $ .101} &  100.0 \% \\
				& $\kappa$ &.720 $ \pm $ .199 & .368 $ \pm $ .198 & .654 $ \pm $ .123 & \textbf{.849 $ \pm $ .113}  & 130.7\%\\%
				\bottomrule
			\end{tabular}
	\end{center}
	\vskip 0.1in}
\end{table}
\endgroup

Table~\ref{tab:benchmark} summarises the F1 and Kappa scores for seven HAR datasets. The random forest models outperformed the deep learning models trained from scratch for all except the \textit{Capture-24} dataset, which is the largest dataset in our evaluations (Table~\ref{tab:dataset}). The performance gap between random forest and training from scratch was the largest in smaller datasets. Meanwhile, pre-trained models outperformed the models trained from scratch and random forest in all seven datasets. Fine-tuning all layers was better than fine-tuning just the fully connected (FC) layers after the ConV layers.

The most significant improvement using pre-training was seen on the small datasets. Conversely, the SSL benefit was more modest for larger datasets. In \textit{Capture-24}, the F1 improvement was 2.5\% when comparing the model with and without SSL pre-training. Nonetheless, with SSL pre-training, the median relative F1 improvement was 18.4\% when compared against the same network trained from scratch and 8.5\% when compared against the random forest model.

\subsection{Transfer learning using labelled pre-training} Even though supervised pre-training can boost the learning outcome substantially than training from scratch (Table~\ref{tab:benchmark} vs Table~\ref{tab:transfer_learning}), it was surprising to see self-supervised pre-training could outperform supervised pre-training when using \textit{Rowlands} and \textit{Capture-24} as the source data. We suspect the limited number of labels that the source datasets had did not contain enough information as what would have been learnt using the SSL pretext tasks.

\begingroup
\setlength{\tabcolsep}{4pt}
\begin{table}[t]
{
    \small
	\caption{Transfer learning (subject-wise F1) performance comparison between supervised pre-training with self-supervised pre-training.} 
	\label{tab:transfer_learning}
	\vskip 0.15in
	\begin{center}
			\begin{tabular}{ c | cc | cc | c}
				\toprule
                 \multirow{3}{*}{Target data} & \multicolumn{5}{c}{Source data} \\
                 \cline{2-6}
                 \\[-.7em]
				 &    \multicolumn{2}{c|}{Rowlands}	 &   \multicolumn{2}{c|}{Capture-24} & UK-Biobank \\
				& Supervised & Self-supervised &  Supervised & Self-supervised & Self-supervised  \\
				\toprule
                Capture-24 & .707 $ \pm $ .094 &  .709 $ \pm $  .094 & - & .707 $ \pm $ .094 &  \textbf{.726 $ \pm $ .093} \\
                Rowlands & - & .734 $ \pm $ .082 & .728 $ \pm $ .094 &	.730 $ \pm $.084 &  \textbf{.796 $ \pm $ .093} \\
				WISDM  &.680 $ \pm $ .109 & .702 $ \pm $  .123& .715 $ \pm $ .119 & .723 $ \pm $  .121 &  \textbf{.810 $ \pm $ .127} \\
				REALWORLD   & .712 $ \pm $ .086 & .737 $ \pm $  .105& .759 $ \pm $ .070 & .771 $ \pm $  .061 & \textbf{.792 $ \pm $ .075}\\
				Opportunity	& .536 $ \pm $ .019 & .539 $ \pm $  .018& .547 $ \pm $ .043 & .547 $ \pm $  .042 & \textbf{.595 $ \pm $ .085} \\
				PAMAP2 & .677 $ \pm $ .082 & .689 $ \pm $  .078& .678 $ \pm $ .118 & .725 $ \pm $  .725 & \textbf{.789 $ \pm $ .054}	 \\	
				ADL  & .634 $ \pm $ .182 & .701 $ \pm $  .111& .768 $ \pm $ .169 & .754 $ \pm $  .159 & \textbf{.829 $ \pm $ .101}  \\
				\bottomrule
			\end{tabular}
	\end{center}
	\vskip 0.1in}
\end{table}
\endgroup

\subsection{Ablation studies}
\begin{figure}[t]
    \begin{minipage}[t]{.495\textwidth}
        \centering
        \includegraphics[width=\textwidth]{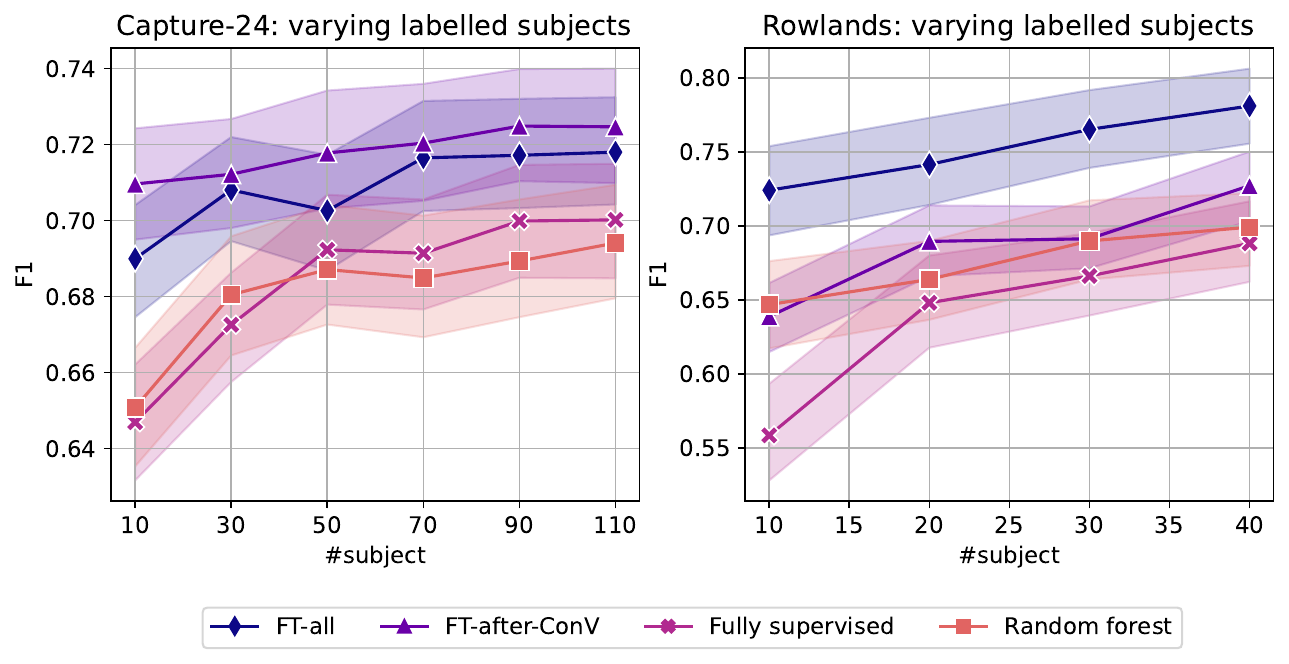}
        \subcaption{How different models perform in the downstream tasks when we change the number of labelled subjects: \textit{Capture-24} (\textbf{left}) and \textit{Rowlands} (\textbf{right}).}\label{fig:varyLsubjects}
    \end{minipage}
    \hfill
    \begin{minipage}[t]{.495\textwidth}
        \centering
         \includegraphics[width=\textwidth]{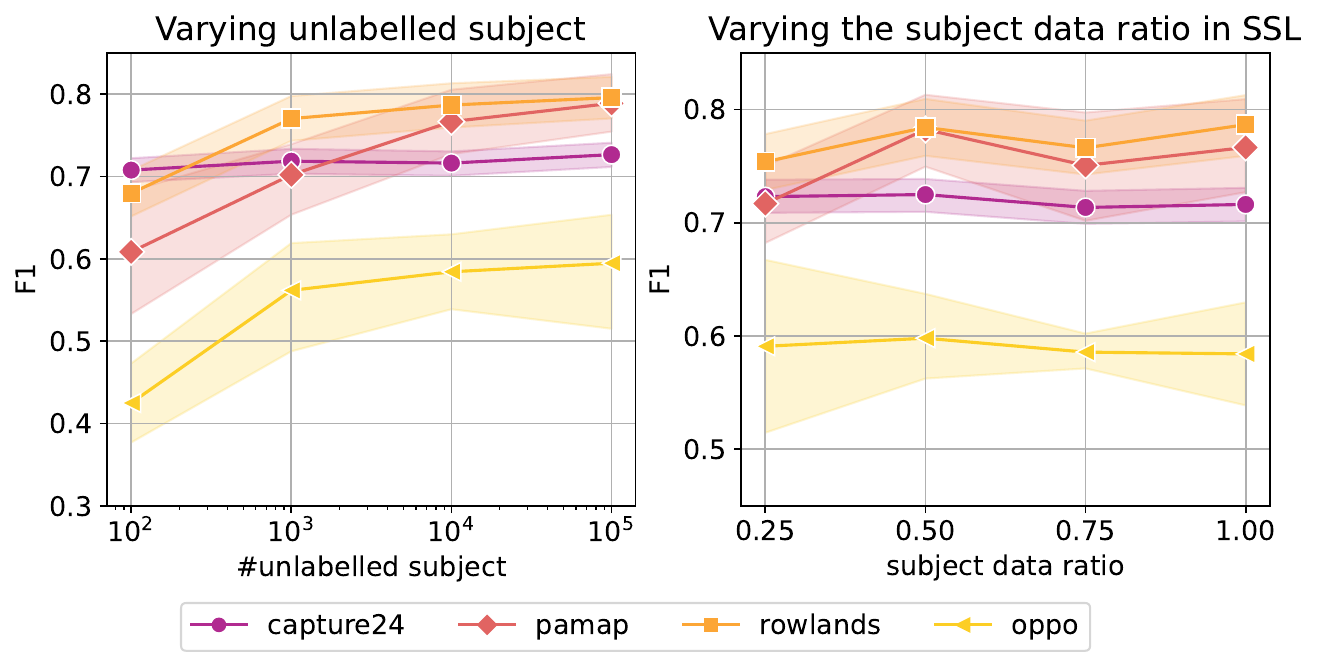}
         \subcaption{\textbf{Left}: More \#unlabelled subjects in the SSL leads to better downstream performance. \textbf{Right}: Unlabelled data per subject does not improve downstream performance when using 10,000 subjects for pre-training.}\label{fig:varySSL}    \end{minipage}  
    \label{fig:1-2}
    \caption{The impact of varying amount of labelled or unlabelled data in the self-supervised training and the fine-tuning stage on subsequent human activity recognition performance. Mean F1$\pm$ SD are plotted.}
\end{figure}

\paragraph{Varying labelled data in the downstream} We observed that pre-trained models did well regardless of the number of labelled subjects in two downstream datasets (Figure~\ref{fig:varyLsubjects}). However, fully-supervised and random forest models were more susceptible to the number of labelled subjects. The performance gain for having more labelled subjects was roughly linear w.r.t. the number of people included with a greater increase when we had fewer labelled subjects.

\paragraph{Varying unlabelled data in the pre-training} We also found that the downstream  HAR performance appeared to increase linearly w.r.t. the number of unlabelled subjects on a log scale ( Figure~\ref{fig:varySSL} left). This conforms with the previous finding in a semi-supervised setting~\citep{oliver2018realistic}. The SSL performance boost with more SSL unlabelled subjects was most significant in the smallest dataset, \textit{Opportunity}. Furthermore, if the number of participants is fixed at 10,000 in pre-training, the data ratio included per subject did not significantly influence the downstream performance. To our surprise, the downstream performance did not degrade too much even when we reduced the subject data ratio from 1.0 to .25 (Figure~\ref{fig:varySSL} right). We suspect that because we had a sufficiently large sample size (10,000), having more data per subject would not help to learn a better representation anymore. The trade-off between inter-subject and intra-subject variability in self-supervision warrants future investigation. Knowing how to prioritize the number of subjects and how much data per subject would benefit the data curation process, especially in fields like medical sciences, where it might be easier to get some data from lots of people than get lots of data from some people.

\subsection{Understanding the representation}
\begin{figure}
    \centering
    \begin{subfigure}[b]{.48\textwidth}
        \includegraphics[width=\textwidth]{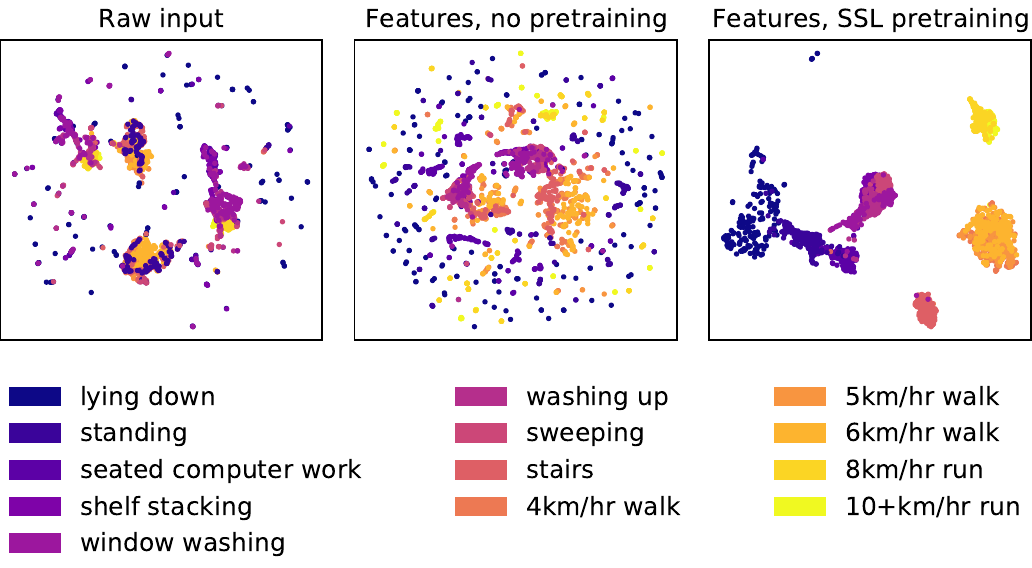}
        \caption{\textit{Rowlands}}
    \end{subfigure}\hfill
    \begin{subfigure}[b]{.48\textwidth}
        \includegraphics[width=\textwidth]{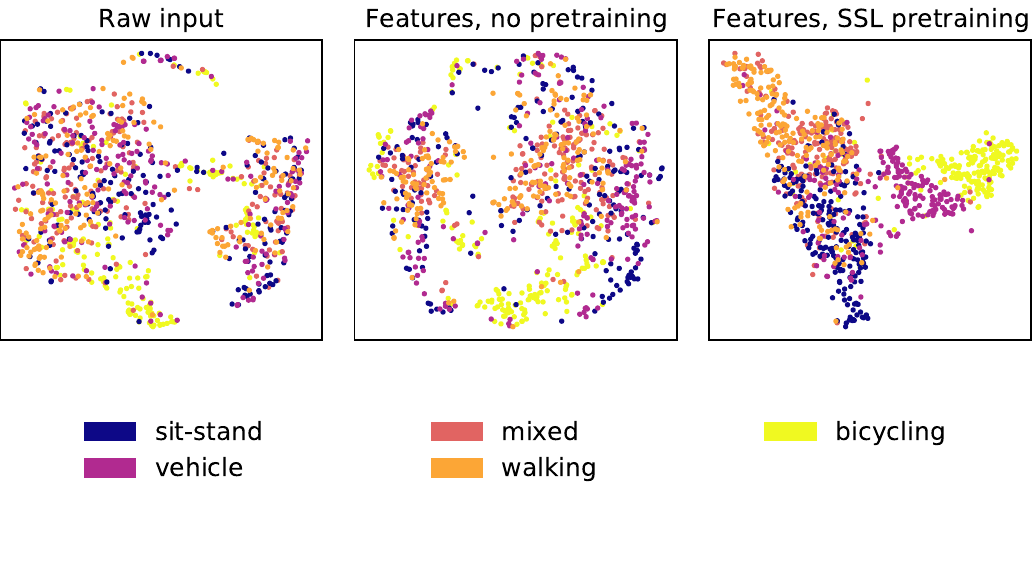}
        \caption{\textit{CAPTURE-24}}
    \end{subfigure}
    \caption{
        Cluster analysis on raw inputs, untrained features and SSL-pretrained features. We use color gradients to denote activity intensities. Results suggest that SSL-derived features are better at clustering similar activities (e.g. walking, stair climbing vs. sitting, writing, typing) as well as their intensities (e.g. lying down, sitting, standing vs. jogging, sports).
    }
    \label{fig:umap}
\end{figure}

 \begin{figure}[t!]
      \centering
      \begin{subfigure}[t]{0.49\textwidth}
          \centering
          \includegraphics[width=\textwidth]{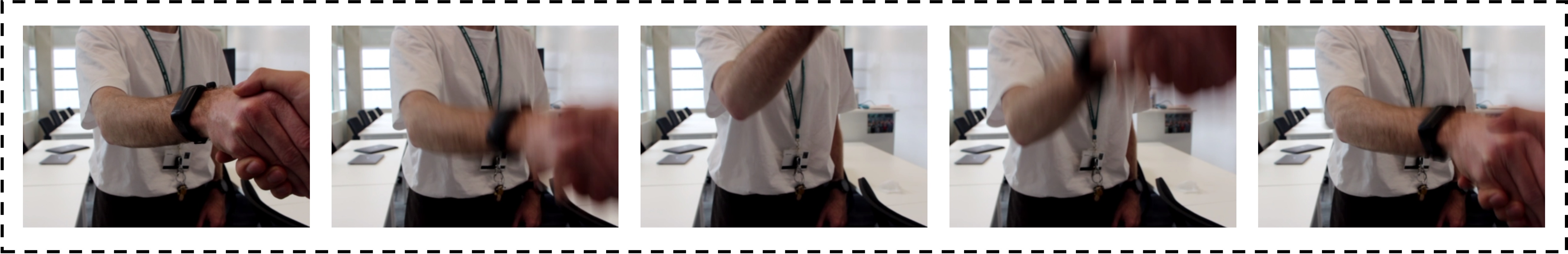}
      \end{subfigure}
        \begin{subfigure}[t]{0.49\textwidth}
         \centering
         \includegraphics[width=\textwidth]{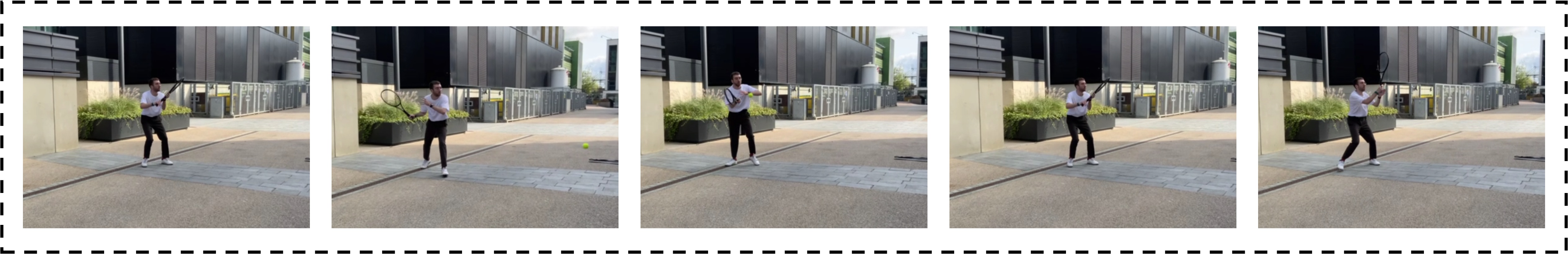}
     \end{subfigure}
      \begin{subfigure}[t]{0.49\textwidth}
          \includegraphics[width=1\textwidth]{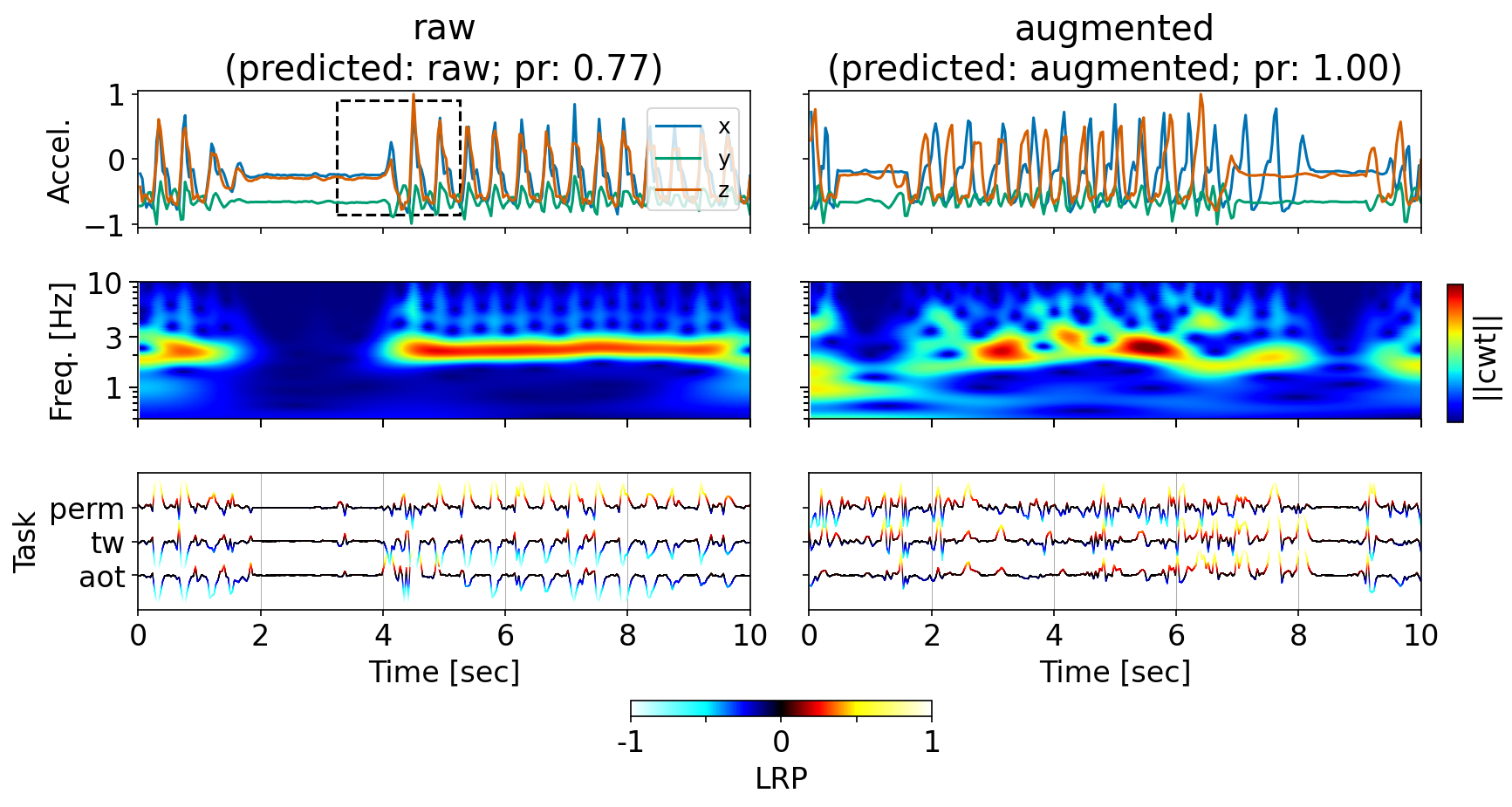}\caption{\footnotesize Shaking hands}\label{fig:lrp-handshake}
      \end{subfigure}
      \begin{subfigure}[t]{0.49\textwidth}
          \includegraphics[width=1\textwidth]{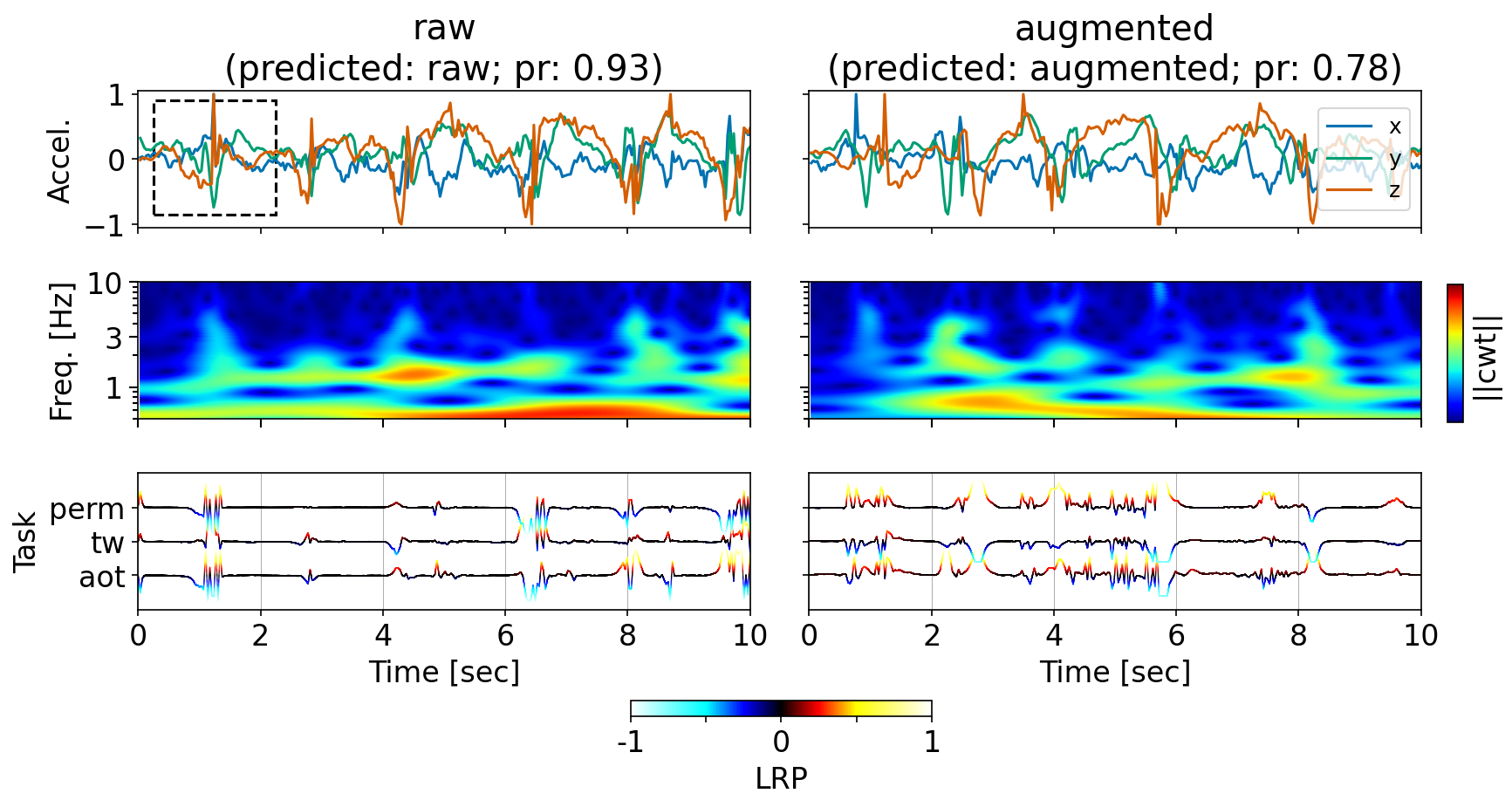}\caption{\footnotesize Playing tennis}\label{fig:lrp-tennis-forehand}
      \end{subfigure}
    \caption{\footnotesize Arrow of time + permutation + time-warped signals of (a) scripted, repetitive low intensity activities, e.g. shaking hands and (b) unscripted, repetitive high intensity activity, e.g. playing tennis. The first row shows the concurrent video frames when a participant performs an activity. The second row is the accelerometry trace. The third row is the continuous wavelet transform scalograms. The last row is the task-specific LRP attribution. Dashed patches over the acceleration trace correspond to the concurrent video video frames. Probability (Pr.) of individual transform applied to: (a) raw: AoT (0.), permutation (.68), TW (0.); augmented: AoT (1.),  permutation (1.), TW (1.); (b) AoT (.03), permutation (.18) , TW (0.) ; augmented: AoT (1.),  permutation (.98), TW (.36); }\label{fig:cLRP}
 \end{figure}

\paragraph{Cluster analysis}
We used UMAP~\citep{mcinnes2018umap} with default parameters for low-dimensional projections for visualization. This was applied to the raw inputs, untrained features, and SSL-derived features \emph{without fine-tuning}. Results for two of the downstream datasets are shown in~Figure~\ref{fig:umap}, and remaining results can be found in Appx.~Figure~\ref{fig:appxumap}. Across all datasets, we observed that the SSL-derived features were better at clustering similar activities (e.g. walking, stair climbing vs sitting, writing, typing) as well as their intensities (e.g. lying down, sitting, standing vs jogging, sports), exhibiting better intra-class compactness and inter-class separability.

\paragraph{Feature interpretation}
Next, we visualised two exemplary pretext SSL task predictions in the presence of repetitive low- and high-intensity activities: shaking hands (Figure~\ref{fig:lrp-handshake}) and playing tennis (Figure~\ref{fig:lrp-tennis-forehand}). During tennis playing, a repetitive high, intensity activity, relevance scores tended to highlight the moments around the natural movements of swinging and hitting the tennis ball (Figure~\ref{fig:lrp-tennis-forehand}, Appx. Figure~\ref{fig:cLRP-appendix}). When performing a repetitive low-intensity activity experiment, for example, shaking hands (Figure~\ref{fig:lrp-handshake}), LRP appeared to also identify the intensity and natural signal periodicity as indicative of the original activity. In contrast, for augmented signals, our model attributed more during periods of visually unrealistic motion dynamics, such as unnatural fragmentation in activity frequency or synchronisation mismatches between sensor axes. Interestingly, stationary movement periods were not relevant for detecting the pretext tasks, further emphasising the importance of our movement-based weighted sampling approach during training. We found that most tasks tended to identify similar regions as relevant. However, this might be expected due to the large shared feature extractor base across all tasks, with smaller MTL heads to identify each pretext task. 

Finally, we empirically compared the faithfulness of the XAI algorithms investigated and the combination of various LRP parameters, using sample-masking experiments for a random subset of 1000 (out-of-sample) subjects in the \textit{UKB} (Appx. Section~\ref{sec:appendix:results:permutation-experiments}). Most XAI models consistently demonstrated the ability to identify relevant patterns for discriminating transformed samples from the original raw data when compared against a random model.

 \section{Conclusion and future work} \label{sec:conclusion}
Our main findings are: 
(1) Self-supervised pre-training consistently improved downstream HAR, especially in the small data regime, reducing the need for labelled data. 
(2) It is possible to learn representations that generalise well across external datasets, tasks, devices and populations -- this contrasts with previous studies where SSL training and evaluation were performed on the same data sources.
(3) Multi-task SSL appears to be learning human motion dynamics, intensity levels, and the synchronisation between different axes (Figure~\ref{fig:umap} and Figure~\ref{fig:cLRP}). Even before fine-tuning, the pre-trained models already have the capacity to distinguish different human activities.

%
In the future, we could include other compatible data modalities such as electrocardiogram data. Having a multi-modal representation would allow us to build foundation models that could be used in critical fields like health research~\citep{bommasani2021opportunities, spathis2022breaking}. Another potential work is the analysis of inter-subject and intra-subject variability in sensor data which could better inform the data curation procedures.

Due to a current lack of raw accelerometer datasets in different regions of the world, a limitation of our work is that the pre-training data (UKB dataset) consists mostly of Caucasians from the UK. A natural extension of this work could include datasets from different demographic groups, as they become available in the future, to improve model fairness.

\medskip

\begin{ack}
We would like to thank all the helpful discussions and feedback we recevied from Aidan Acquah, Gert Mertes, Henrique Aguiar, Andres Tamm, and Korsuk Sirinukunwattana.

This research has been conducted using the UK Biobank Resource under Application Number 59070. This work is supported by: Novo Nordisk (HY, AD); the Wellcome Trust [223100/Z/21/Z] (AD); GlaxoSmithKline (AC, DC); the British Heart Foundation Centre of Research Excellence [RE/18/3/34214] (AD); the National Institute for Health Research (NIHR) Oxford Biomedical Research Centre (AD, DC); and Health Data Research UK, an initiative funded by UK Research and Innovation, Department of Health and Social Care (England) and the devolved administrations, and leading medical research charities. It is also supported by the UK’s Engineering and Physical Sciences Research Council (EPSRC) with grants EP/S001530/1 (the MOA project) and EP/R018677/1 (the OPERA project); and the European Research Council (ERC) via the REDIAL project (Grant Agreement ID: 805194), and industrial funding from Samsung AI.

We would also like to thank Alex Rowlands and Mike Catt, who kindly shared their activity dataset with us. Their project was funded by a grant from Unilever Discover to the School of Sports and Health Sciences, University of Exeter.

No funding bodies had any role in the analysis, decision to publish, or preparation of the manuscript.
\end{ack}

\bibliography{ssl}
\bibliographystyle{apalike}

\clearpage

\appendix

\section{Appendix}

\subsection{Datasets}

\begin{table}[ht]
{
	\caption{The license, consent and device used for each dataset}
	\label{tab:extraInfo}
	\vskip 0.15in
	\begin{center}
			\begin{tabular}{ccp{3cm} p{3cm} p{2cm}}
				\toprule
				Dataset &  Personal Info  & Licence & Consent & Device  \\
				\midrule
				UK-Biobank &  \xmark & Non-exclusive licence (but not ownership rights) for permitted purpose only  &  Informed consent for research  &  Axitivity (AX3) \\
				Capture-24 &   \xmark & CC Attribution & Informed consent for research  & Axitivity (AX3) \\
				Rowlands &     \xmark & Non-specified &  Informed consent for research  & Genea \\
				WISDM    & \xmark  & Non-specified  & Non-specified &  LG G Watch \\
                REALWORLD    &   \xmark & Non-specified & Non-specified & LG G Watch R   \\
            	Opportunity & \xmark &  Non-specified & Non-specified  & Custom inertial measurement units (IMUs)  \\
				PAMAP2    & \xmark & Non-specified  &  Informed consent for research & Colibri inertial measurement units (IMUs)  \\
				ADL & \xmark  & Free for research & Non-specified & Non-specified \\
				\bottomrule
			\end{tabular}
	\end{center}
	\vskip -0.1in
}
\end{table}

\subsection{Methods}
\subsection{Feature engineering} \label{appx:feature}
The list of hand-crafted features that were extracted for random forest:
\begin{itemize}
    \item Mean, standard deviation, and range for each axis.
    \item Correlation between each axis pair.
    \item Euclidean norm, its mean, standard deviation, range, median absolute deviation, kurtosis, and skew.
    \item The top two dominant frequencies in the power spectrum.
\end{itemize}

\subsubsection{Explainable AI Framework for Time-Series Prediction}\label{sec:appendix:methods:xai}
Holistic visual interpretation of the AoT predictions were determined by (1) visualising the raw data, (2) its time-frequency representation using the (discretised) continuous wavelet transform (CWT) and (3) assessing the time-localised LRP attribution.
\paragraph{The Continuous Wavelet Transform:} The CWT is a method used to measure the similarity between a signal and an analysing function (in this case the Morlet wavelet) which can provide a precise time-frequency representation of a signal \citep{RN988, RN701}. 
\paragraph{Layer Wise Relevance Propagation:}
The LRP algorithm back-propagates through a network to decompose the final output decision, $ f(\bm{x})$ \citep{samek2021explaining, montavon2019lrp}. Briefly, a trained model's activations, weights and biases are first obtained in a forward pass through the network. Secondly, during a backwards pass through the model, LRP attributes relevance to the individual input nodes, layer by layer. For example $R_{k}$ denotes the relevance for neuron $k$ in layer $^{(l+1)}$, and $R_{j\leftarrow k}$ defines the share of $R_{k}$ that is redistributed to neuron $j$ in layer $^{(l)}$. The fundamental concept underpinning LRP is that the conservation of relevance per layer, which can be denoted as:  
\begin{equation}\label{eq:conserve_relevance}
\sum_{j}R_{j\leftarrow k}^{(l)} = R_{k}^{(l+1)}
\end{equation}
The LRP algorithm initiates at the model output and iterates over all layers in the model in a backwards pass until the relevance scores $R_i$ for all inputs of $x_i$ are computed. Relevance values $R_i > 0$ signify components $x_i$ which represent the presence of the predicted class, while conversely $R_i < 0$ contradict the prediction of that class. $R_i \approx 0$ indicate inputs $x_i$ which have little or no influence to the model’s decision. 

It has been demonstrated that the combination of different rules throughout a network yields the most faithful and understandable LRP explanations \citep{kohlbrenner2020towards}. As such, we applied a composite rule (LRP-CMP) with LRP-$\gamma$ applied to the shallower 
convolutional layers, LRP-$\epsilon$ rules ($\epsilon$=$\{1e^{-9}, 1e^{-3}, 10\}$) in the middle of the network, and LRP-$0$ applied to the final linear classification layer. 

\paragraph{Visually interpreting SSL attribution:}
To visually interpret SSL attribution, we compared the raw time-series accelerometery with the analogous SSL task transformed data. In in the panel plot depicted in figures \ref{fig:cLRP} and \ref{fig:cLRP-appendix}, the top rows represent the 3-axis accelerometer trace for each channel: $(\textit{\textbf{a}}_x, \textit{\textbf{a}}_y, \textit{\textbf{a}}_z)$; the second rows depict the top view of the continuous wavelet transform (CWT) scalogram of $\|\textit{\textbf{a}}\|$, which is the absolute value of the CWT as a function of time and frequency. The bottom rows denote the relevance values ($R_i$) attributed using LRP. Red and hot colours identify input segments where $R_i>0$ (contribution to a class prediction), whereas blue and cold hues identify $R_i<0$ (contradicting a class prediction), while black represents ($R_i\approx0$) inputs which have little or no influence to the model's decision. Square patches over the raw accelerometer trace correspond to the video frames depicted above each panel plot.

\paragraph{Evaluating XAI algorithm faithfulness:} 
In order to test the faithfulness of an explanation provided by an XAI framework, a sample-masking experiment was performed, comparing some popular XAI models and LRP parameters options. Briefly, to conduct a permutation analysis, the most relevant samples identified from each XAI algorithm were cumulatively masked (imputation with random Gaussian noise), from most relevant to least relevant \citep{samek2021explaining}. As such, the faster the accuracy of the model decreases with the number of masked samples, the more faithful the explanation method is with respect to the decision of the neural network. 

Permutation tests were performed in random batches containing correctly identified duplicate original and augmented samples (AoT + permutation + TW) from 1000 subjects in the out-of-sample test set sampled in the \textit{UKB}. The mean degradation in SSL prediction through our permutation experiment is shown in figure \ref{fig:permutation_test}. We compared various LRP parameter options, LRP-$0$, LRP-$\epsilon$, LRP-CMP, as well as some popular off-the-shelf XAI attribution frameworks, such as saliency mapping \citep{simonyan2013deep}, Guided Backpropagation (GBP) \citep{springenberg2014striving} and Integrated Gradients (IG) \citep{sundararajan2017axiomatic}.

\clearpage

\subsection{Results}
\begin{figure}[h]
    \centering
    \includegraphics[width=0.5\linewidth]{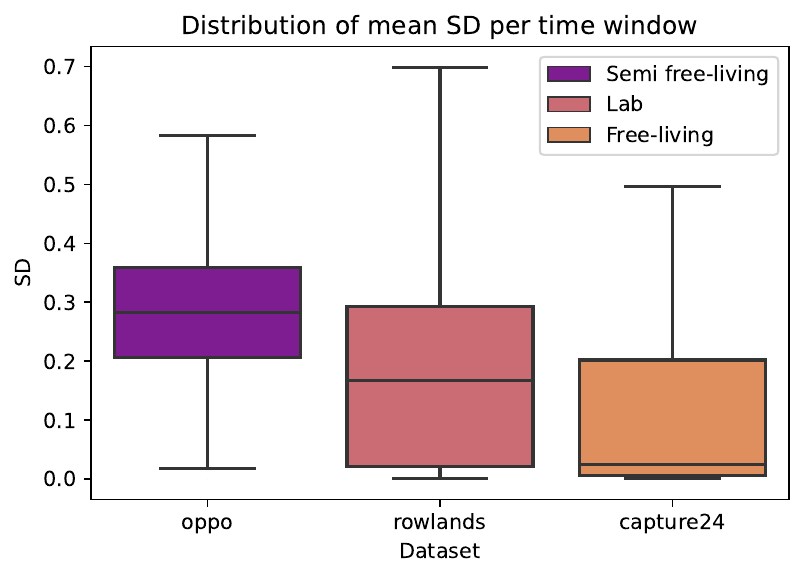}
    \caption{\footnotesize Data collected from free-living environment tends to have more stationary periods than the data collected in lab environment.}
    \label{fig:sd}
\end{figure}

\begin{table}[ht]
{\scriptsize
	\caption{Downstream human activity recognition performance (subject-wise F1 ($\pm$SD)) when using different self-supervised learning tasks after fine-tuning: Arrow of the Time (AoT), Permutation (P), Time warping (TW). The tasks were trained using the same 1,000 UK-Biobank participants.} 
	\label{tab:single_task}
	\vskip 0.15in
	\begin{center}
			\begin{tabular}{lccccccc}
				\toprule
				Task &  Capture-24 & Rowlands & WISDM  & REALWORLD & Opportunity & PAMAP2   & ADL \\
                \#Subjects & 152 & 55 & 46 & 14 & 4 & 8 & 7\\
				\#Samples & 573K &   36K & 28k & 12k &  3882 &  2869 & 635 \\
				\toprule
                AoT 	 &  .671 $ \pm $ .094 & .565 $ \pm $ .120 & .767 $ \pm $ .124 & .750 $ \pm $ .084 & .582 $ \pm $ .054 & .715 $ \pm $ .036  & .754 $ \pm $ .157 \\
 
			    Permutation  & \textbf{.721 $ \pm $ .093} & \textbf{.783 $ \pm $ .099} & \textbf{.778 $ \pm $ .109} & .766 $ \pm $ .063 & \textbf{.588 $ \pm $ .076} & \textbf{.750 $ \pm $ .057} & \textbf{.810 $ \pm $ .097} \\%
			    
				TW  	& .715 $ \pm $ .093 & .776 $ \pm $ .110 & .767 $ \pm $ .120 & \textbf{.772 $ \pm $ .073} & .584 $ \pm $ .064 & .737 $ \pm $ .079 & .765 $ \pm $ .117\\%
				\bottomrule
			\end{tabular}
	\end{center}
	\vskip -0.1in
}
\end{table}

\begin{table}[ht]
{\small
	\caption{Test accuracies on different tasks 1000 subjects.} 
	\label{tab:acc_tasks}
	\vskip 0.15in
	\begin{center}
			\begin{tabular}{cccc}
				\toprule
				AoT   &  Permutation & Time warped \\
				\toprule
                0.93   &  .90     &  .97  \\
				\bottomrule
			\end{tabular}
	\end{center}
	\vskip -0.1in
}
\end{table}

\clearpage

\subsubsection{Additional UMAP visualizations}
\begin{figure}[ht]
    \centering
        
    \begin{subfigure}[b]{.48\textwidth}
        \includegraphics[width=\textwidth]{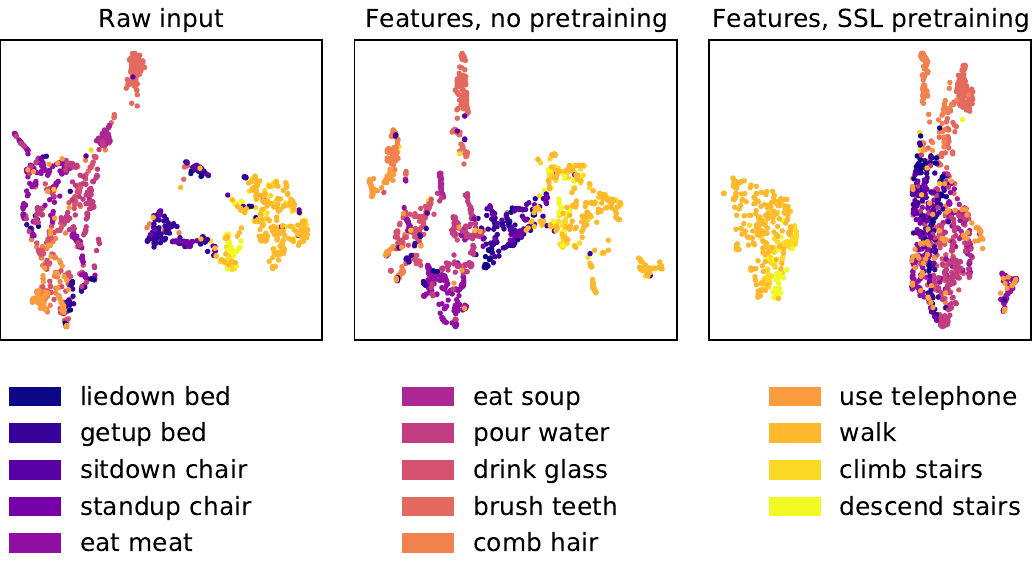}
        \caption{\textit{ADL}}
    \end{subfigure}\hfill
    \begin{subfigure}[b]{.48\textwidth}
        \includegraphics[width=\textwidth]{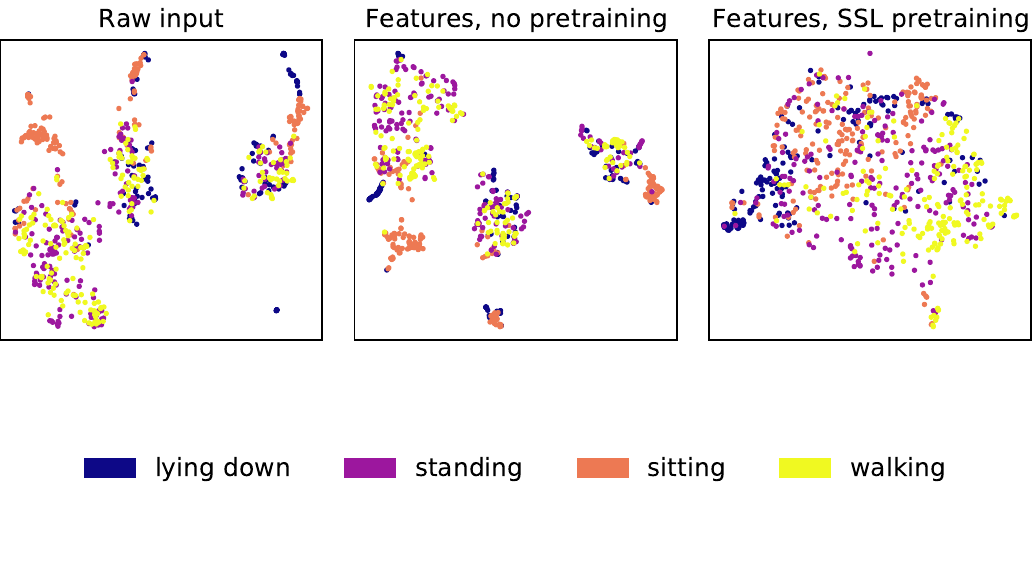}
        \caption{\textit{Opportunity}}
    \end{subfigure}
    \vspace{1.5em}

    \begin{subfigure}[b]{.48\textwidth}
        \includegraphics[width=\textwidth]{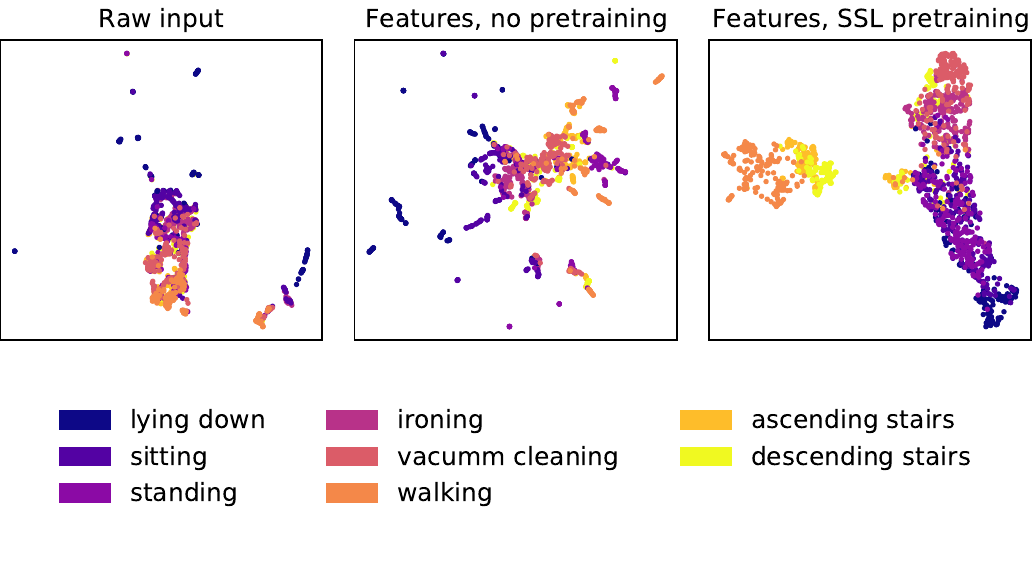}
        \caption{\textit{PAMAP2}}
    \end{subfigure}\hfill
    \begin{subfigure}[b]{.48\textwidth}
        \includegraphics[width=\textwidth]{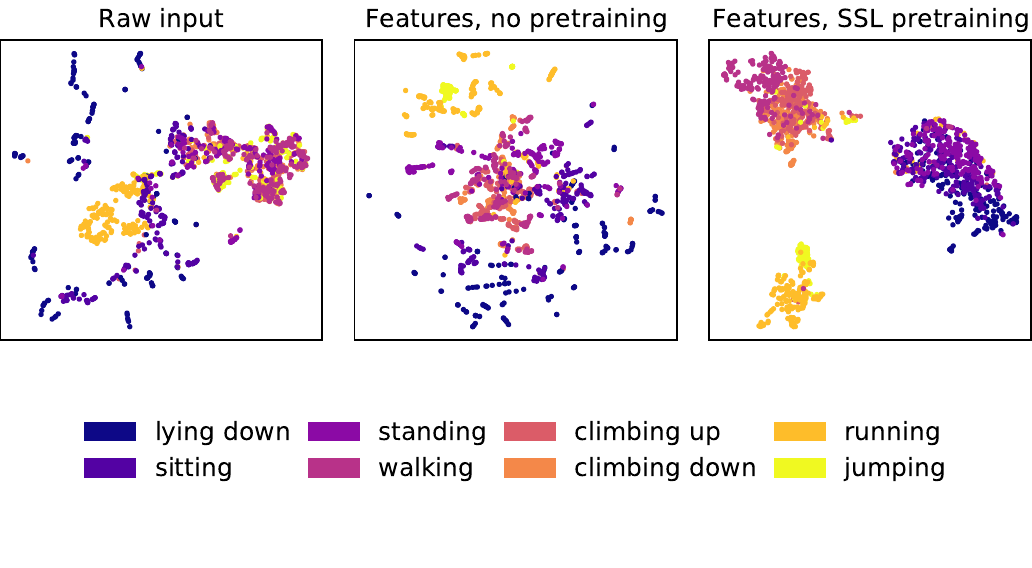}
        \caption{\textit{REALWORLD}}
    \end{subfigure}
    \vspace{1.5em}
    
    \begin{subfigure}[b]{.48\textwidth}
        \includegraphics[width=\textwidth]{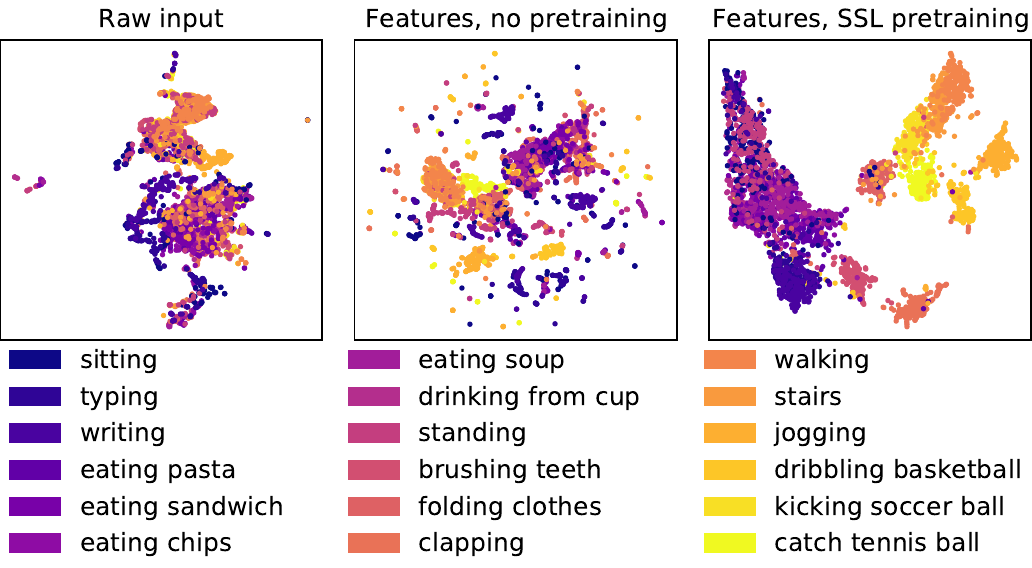}
        \caption{\textit{WISDM}}
    \end{subfigure}

    \caption{
        Cluster analysis on raw inputs, untrained features and SSL-pretrained features. We use color gradients to denote activity intensities. Results suggest that SSL-derived features are better at clustering similar activities (e.g. walking, stair climbing vs. sitting, writing, typing) as well as their intensities (e.g. lying down, sitting, standing vs. jogging, sports).
    }
    \label{fig:appxumap}
\end{figure}

\clearpage

\subsubsection{Explainable Framework Performance}\label{sec:appendix:results:permutation-experiments}
\begin{figure}[t!]
    \centering
    \includegraphics[width=0.75\linewidth]{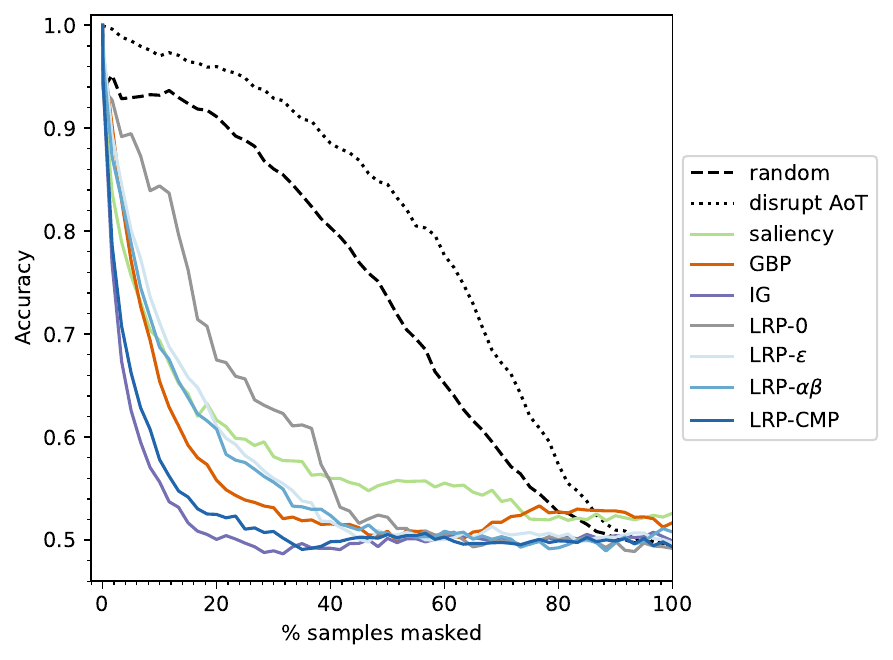}
    \caption{\footnotesize Comparison of popular explainability algorithm performance on AoT + permutation + TW for 1000 random subjects in the UK Biobank out-of-sample test data, through consecutively masking samples from most relevant to least relevant. We observed that, for most explanation methods, removing identified relevant features quickly destroyed class evidence, thus degrading the out-of-sample test accuracy of the model compared to randomly masking samples.}
    \label{fig:permutation_test}
\end{figure}

We observed that, for most explanation methods, removing relevant features quickly destroyed class evidence, degrading the out-of-sample test accuracy of the SSL model, as shown in figure \ref{fig:permutation_test}. This inferred that the XAI methods had fatefully identified relevant patterns within the accelerometer signal for determining the SSL pretext task compared to randomly mixing and masking samples in the signal. Adding random noise to the accelerometer was demonstrated to degrade model performance (which is to be expected) but at a slower rate than masking the most relevant samples first. 

Furthermore, replacing the accelerometer signal with noise, cumulatively from the first time-step until the last (essentially disrupting the AoT), was found to degrade AoT-specific model accuracy at a slower rate than masking samples based on XAI relevance sorting or randomly. Note: AoT disruption was calculated from the start of the signal to the end (forward) and from the end of the signal to the start (reverse) and presented as the mean AoT disruption. As such, the the model does not appear to learn a single sequential pattern within a time-series (either forward or reverse), but instead suggests that there may be certain signal morphologies that may indicate the SSL augmentation, such as the characteristics of dynamic human motion.

\subsubsection{Further Contextual LRP Examples}\label{sec:appendix:results:contextual-lrp}
\begin{figure}[th!]
      \centering
      \begin{subfigure}[t]{0.49\textwidth}
          \centering
          \includegraphics[width=\textwidth]{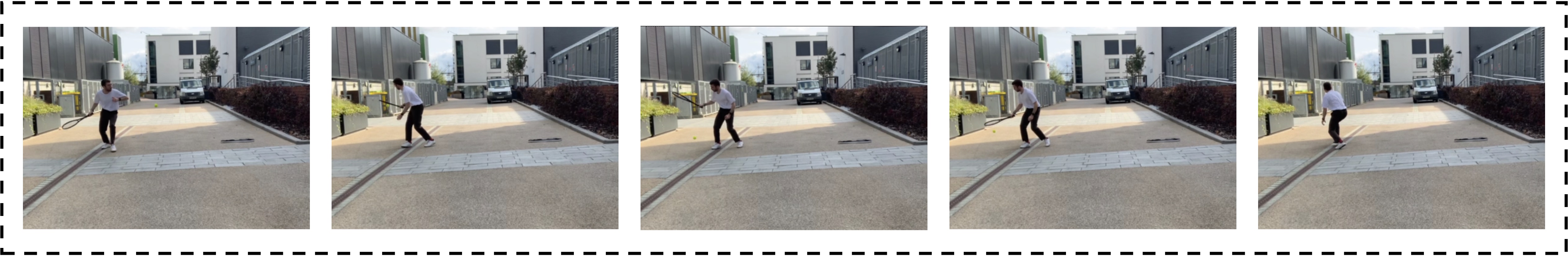}
     \end{subfigure}
      \begin{subfigure}[t]{0.49\textwidth}
          \centering
          \includegraphics[width=\textwidth]{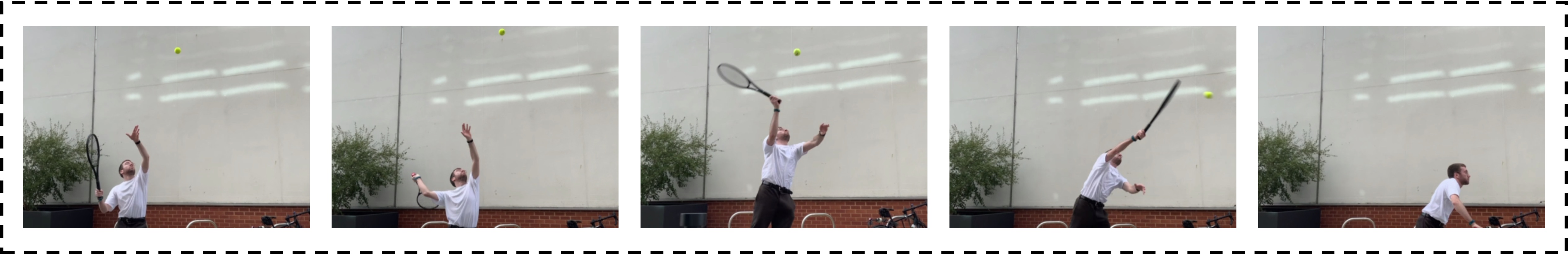}
   \end{subfigure}
      \begin{subfigure}[t]{0.49\textwidth}
          \centering
          \includegraphics[width=\textwidth]{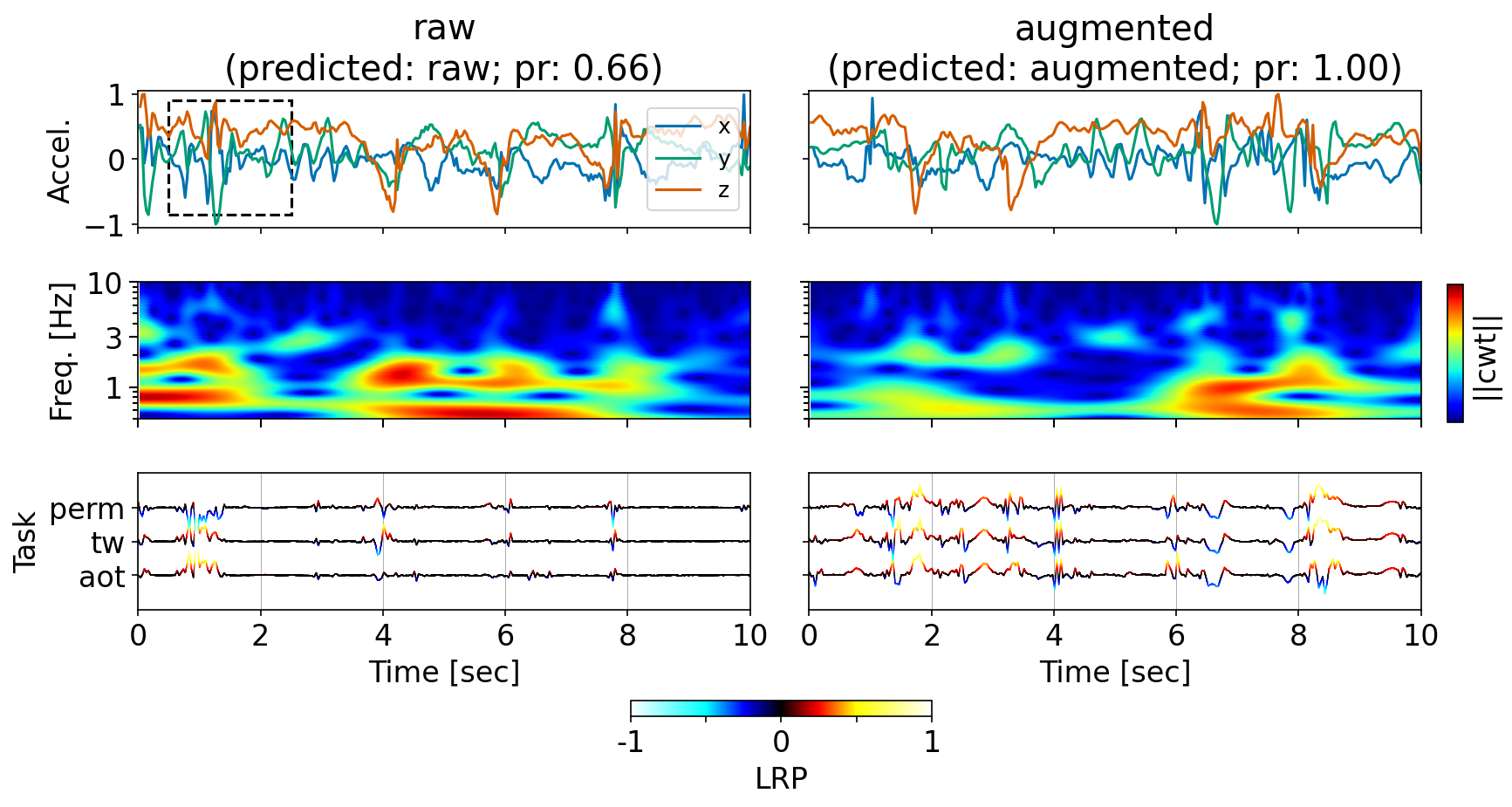}
          \caption{\footnotesize playing tennis (unstructured dropping the ball) }\label{fig:lrp-tennis-backhand}
      \end{subfigure}
    \begin{subfigure}[t]{0.49\textwidth}
         \centering
          \includegraphics[width=\textwidth]{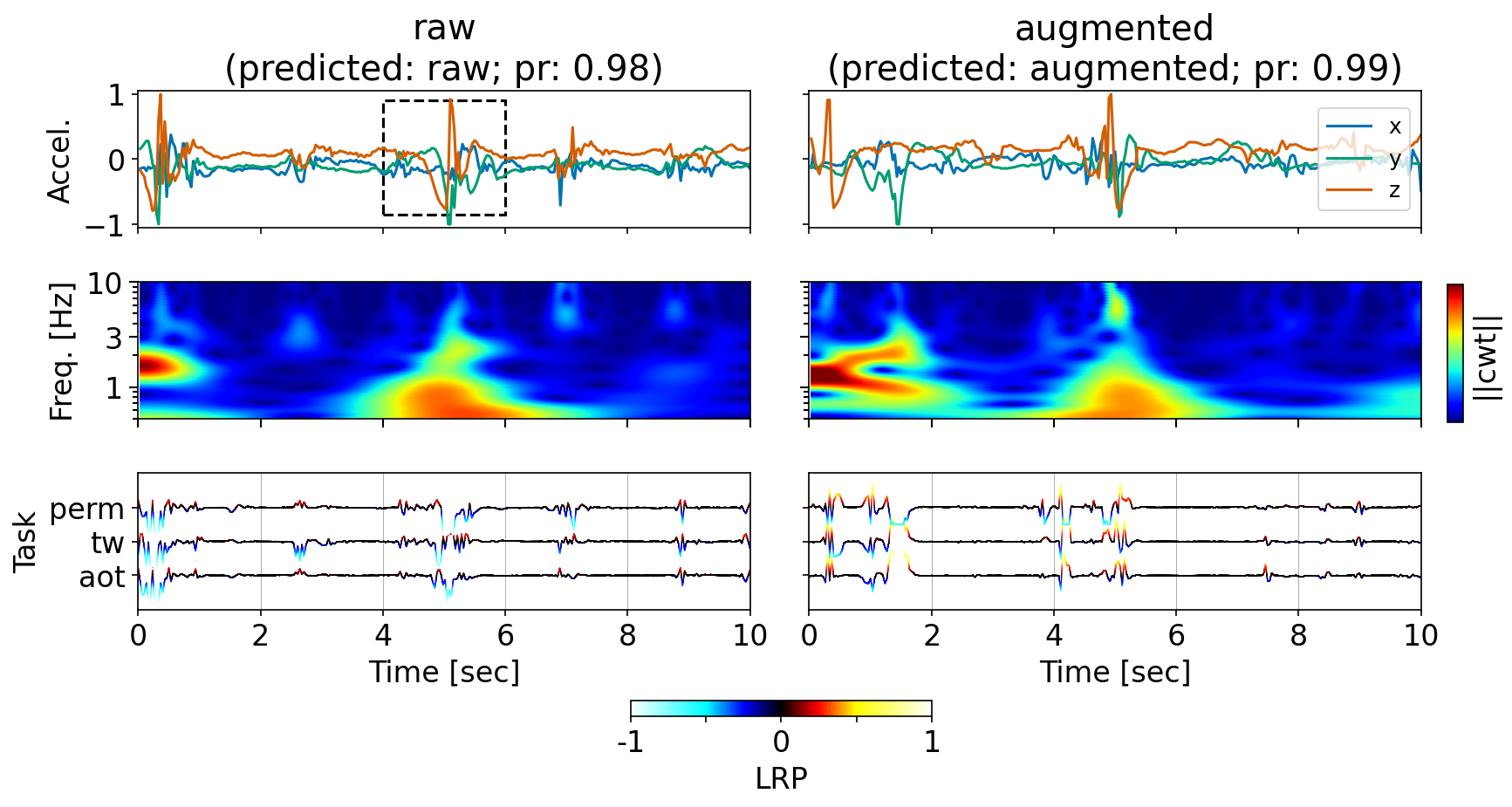}\caption{\footnotesize playing tennis (serve) }\label{fig:lrp-tennis-serve}
      \end{subfigure}
      \caption{\footnotesize 
      Arrow of time + permutation + time-warped signals during unscripted, repetitive high intensity intensity activity of playing tennis. The square patches over the acceleration trace correspond to the video frame depicted during when the participant (a) dropped the ball and (b) performed a tennis serve. Probability (Pr.) of individual transform applied to: (a) raw: AoT (.01), TW (0.), perm (1.); augmented: AoT (1.), TW (1.), perm (1.); (b) raw: AoT (.01), TW (0.), perm (.04); augmented: AoT (1.), TW (1.), perm (.98);}\label{fig:cLRP-appendix}
           
 \end{figure}

Figures \ref{fig:lrp-tennis-backhand} and \ref{fig:lrp-tennis-serve} depict further examples of contextual LRP collected during the during unscripted, repetitive, high intensity activity of playing tennis. For example, figure \ref{fig:lrp-tennis-backhand} depicts when the participant dropped the ball; figure \ref{fig:lrp-tennis-serve} illustrates an overhead tennis serve. It was observed that LRP tended to also attribute relevance to the moments of natural human motion captured during these shots.

\end{document}